\documentstyle[12pt]{article}

\renewcommand{\thefootnote}{\fnsymbol{footnote}}
\newcommand{\EQ}{\begin{equation}}
\newcommand{\EN}{\end{equation}}
\newcommand{\bea}{\begin{eqnarray}}
\newcommand{\ena}{\end{eqnarray}}
\newcommand{\vs}[1]{\vspace{#1 mm}}

\newcommand{\uda}{\nearrow \kern-1em \searrow}

  \newcommand{\iao}{ia\omega}
\newcommand{\G}{\Gamma}

\begin{document}

\topmargin 0pt
\oddsidemargin 5mm

\begin{titlepage}
\setcounter{page}{0}
\begin{flushright}
OU-HET 221\\
August, 1995
\end{flushright}

\vs{15}
\begin{center}
{\Large Absorption probability of De Sitter horizon for massless fields with
spin} \vs{15}

{\large
Hisao Suzuki\footnote{e-mail address: suzuki@phys.wani.osaka-u.ac.jp} and
Eiichi Takasugi\footnote{e-mail address: takasugi@phys.wani.osaka-u.ac.jp}}
\\ \vs{8}
{\em Department of Physics, \\
Osaka University \\ Toyonaka, Osaka 560, Japan} \\ \end{center}
\vs{10}

\centerline{{\bf{Abstract}}}
The evaluation of the absorption coefficients are important for particle
emission caused by Hawking radiation. In the case of cosmological particle
emission from the event horizon in De Sitter space, it is known that the
scalar wave functions are solved in terms of Legendre functions. For fields
with higher spin, the solution has been examined with low frequency
approximation.
We shows that the radial equations of the fields with spin
$0,1/2,1$ and $2$ can be solved analytically in terms of the hypergeometric
functions.
We calculate the absorption probability using asymptotic expansion for high
frequency limit. It turns out that the absorption coefficients are universal
to all bosonic fields; They depend only on the angular momentum and not
spin. In the case of spin $1/2$ fermions, we can also find non-vanishing
absorption probability in contrast to the previously known result.

\end{titlepage}
\newpage
\renewcommand{\thefootnote}{\arabic{footnote}} \setcounter{footnote}{0}

Since the celebrated discovery of the evaporation of the black
holes\cite{Hawking} much effort has been given to get analytic expression of
the
evaporation rate\cite{Page}. The fact that the the corresponding field
equations around black holes can not be expressed by the special functions
prevent us from getting analytic expression for high-frequency mode. On the
other hand, the Hawking radiation caused by the cosmological event
horizon\cite{GH}
seems to be much easier to handle because it is known that the scalar field
equation in static coordinates of De Sitter space can be solved by Legendre
functions\cite{RPone}. We can
therefore convert the solution at the event horizon to our observing world
by analytic continuation for all the frequency. Also the Bogoliubov
coefficients of the mode functions from the global coordinates has also
been obtained\cite{SS}. As for the fields with spin, the solution was made
only by low frequency approximation\cite{RPtwo} where they did not found
any radiation for spinor fields. Our aim of this short letter is to show that
the the radial equation for spin fields can also be solved by special
functions. We obtain the absorption probability using high frequency
approximation. It turns out that the absorption probability for bosonic fields
are universal and do not depend on the spin. For spinor, we find non-vanishing
result for the absorption probability in contrast to the result given in
ref.\cite{RPtwo}.

 De Sitter metric in static coordinates takes the form; \bea
ds^2 = [1-({r^2 \over a^2})]dt^2 - [1-({r^2 \over a^2})]^{-1}dr^2 -
r^2(d\theta^2
+ \sin\theta^2 d \varphi^2), \ena
where a is related to the cosmological constant by $\Lambda = 3/a^2$. In the
case of the type D space-time with vanishing electromagnetic fields, the
equation of the first order variation of the scalar, spin $1/2$ fermions,
electromagnetic fields and gravity is universal and written by Teukolsky
equations\cite{Teukolsky} and separable. In our case, the angular component can
be solved in
terms of the spin-weighted spherical harmonics ${}_sY^m_l({\theta})$ given
in Ref.\cite{Goldberg}. Then the radial equation is given in the variable $z
\equiv r/a$ as\cite{RPtwo} \bea
&{}&z^2(1-z^2)^2({d \over dz})^2 R_p - [2(p+1)z^3 -2(s+1)(1-z^2)z](1-z^2){d
\over dz}R_p \nonumber\\ &{}& - \lbrace{
(1-z^2)[(l-s)(l+s+1)+(s+p+1)(s+p+2)z^2]-(a\omega z)^2 + 2ia\omega z p
\rbrace}R_p\nonumber\\ &=&0 ,
\ena
where $p$ takes the value $\pm s$.
In the case of the scalar fields $s=p=0$, by choosing variable as $y = z^2$,
the singular point of the equation reduces to $0,1$ and $\infty$ with
definite singularity. The equation can then be solved in  hypergeometric
functions\cite{RPone}. This choice of the variable is a natural choice because
the metric
depends only on $z^2$. For $s \neq 0$, however, this choice does not work.
There is
another natural choice of variable. Near horizon, a useful coordinate is given
by $dz^* = dz/(1-z^2)$ and therefore we choose a variable as \bea
y = e^{-2z^*} = {1-z \over 1+z}.
\ena
With this choice of the variable, Eq.(2) can be written as
\bea
&{}& ({d \over dy})^2 R_p + [(p+1){1 \over y} - 2(s+1){ 1 \over 1-y}
-2(s+p+1){1 \over y+1}]{d \over dy} R_p \nonumber\\ &-& \lbrace (l-s)(l+s+1){1
\over (1-y)^2} - (s+p+1)(s+p+2){1 \over (y+1)^2}+ [-{(a\omega)^2 \over 4} + {
ia \omega p \over 2 } ] \nonumber\\ &+&[-(l-s)(l+s+1)-ia\omega p]{1 \over
y(y-1)}+ (s+p+1)(s+p+2){1 \over y(y+1)}\rbrace R_p = 0.
\ena
Now the equation have four defenite singularity at points $0,-1,+1$ and
$\infty$. It turns out that the singularity at $y=-1$ can be factorized as
\bea
R_p=y^{-p-{\iao \over 2}}(1-y)^{l-s}(1+y)^{s+p+1}f_p
\ena
and $f_p$ satisfies the hypergeometric equation
\bea
y(1-y)f_p'' +[1-p-\iao -(2l+3-p-\iao)y]f_p'-(l+1-p)(l+1-\iao)f_p =0,
\ena
which has two independent solutions. Out of two solutions, we take the
outgoing solution at the future horizon $y\sim 0 (z \sim 1)$ which  behaves
as $\exp (2p+\iao)z^*$ where
$z^*=-(\log y)/2$. The solution is
 \bea
R_p = y^{-{ia\omega \over 2}-p}(1-y)^{l-s}(y+1)^{s+p+1}F(a,b;c;y) ,
\ena
where $F(a,b;c;y)$ is the hypergeometric function with
\bea
a=l+1-p,\qquad b=l+1-\iao, \qquad c=1-p-\iao .
\ena

By using the recursion relation of the hypergeometric functions, we can
derive the recursion relations which relate the solutions with different
spins;
\bea
&&{d \over dy}[y^{-{\iao \over 2}}(1-y)^{2s+1}(1+y)^{-2s-1}R_s] \nonumber\\
&& \hskip 1cm
= - (s+\iao)y^{-{\iao \over 2}}(1-y)^{2s+1}(1+y)^{-2s-3}R_{s+1},\\
&&{d \over dy}[y^{{\iao \over 2}+s+1}(1-y)(1+y)^{-2s-3}R_{s+1}] \nonumber\\
&&\hskip 1cm
= - {(l-s)(l+s+1) \over s+\iao }
y^{{\iao \over 2 +s}}(1-y)^{-1}(1+y)^{-2s-1}R_{s}.
\ena
We can also obtain the operations which change the angular momentum.

The behavior of this function near us $(y \simeq 1, z \sim 0)$ can be obtained
by the
analytic continuation. Near $z=0$, the variable $y$ can be written as $y
\simeq e^{-2z},\quad 1-y \simeq 2z$. Since the parameters in the hypergeometric
function satisfies a relation $c-a-b = - (2l+1)$, the analytic
continuation of the function produces logarithmic terms and the analysis
becomes very complicated. Therefore, we will analytically continue the
angular momentum to avoid such complication.(Strictly speaking, we cannot
justify this procedure, although this technique is also used in the case of
Black hole\cite{Page}.) Around $z=0$, $R_p$ behaves as
\bea
R_p &{}&=
(\frac{1-z}{1+z})^{-p-\frac{\iao}2}(\frac{2z}{1+z})^{l-s}(\frac2{1+z})^{s+p+1}\nonumber \\
&& [ {{\G(1-p-\iao) \G (-(2l+1))} \over {\G(-l-\iao)\G(-l-p)}}
   F(l+1-p,l+1-\iao;2l+2;\frac{2z}{1+z}) \nonumber\\
&&+ {{\G(1-p-\iao)\G(2l+1) }\over
\G(l-p+1)\G(l+1-\iao)}(\frac{2z}{1+z})^{-(2l+1)}
F(-l-p, -l-\iao ;-2l;\frac{2z}{1+z})].
\ena
We are going to consider the  high frequency approximation $a\omega >>1$.
By using a formula
\bea
F(a,b;c;x) \sim F(a;c;bx),
\ena
when $b \rightarrow \infty$, and also considering the region where  $\vert
awz \vert >> 1$
and $\mid z \mid <<1$. Then we can use an asymptotic expansion of the
confluent hypergeometric functions;
\bea
F(a;c;x) \sim {\G(c) \over \G(c-a)}\Bigl({-1 \over x}\Bigr)^a
+ {\G(c) \over \G(a)}e^x x^{a-c}
\ena
and find  the following asymptotic expansion of $R_p$;

\hskip 6mm
\bea
R_p &\sim&(-1)^{l+p}{\G(1-p-\iao)\over \G(l+1-\iao)}\nonumber \\
&&\{\frac{e^{\iao z}}{ z^{s-p+1}}4^p
(\iao )^{l+p} [1 - (-1)^{2p}(\iao)^{-(2l+1)}{\G(l+1-\iao) \over
\G(-l-\iao)}]\nonumber\\
&&\quad +\frac{e^{-\iao z}}{ z^{s+p+1}}(-\iao)^{l-p}{\G(-l+p) \over \G(-l-p)}
 [1 + (-1)^{2l}(\iao)^{-(2l+1)}{\G(l+1-\iao) \over \G(-l-\iao)}] \}.
\ena
We observe that $R_p$ is not singular in the limit $2l=$integer.
In order to calculate the absorption coefficients, we use a trick used in
Refs.\cite{TeukolskyPress,Page,RPtwo}. When we write the asymptotic
expansion of $R_s$ as
\bea
R_s
\sim Y^s_{in}e^{-\iao z}/z^{1+2s} + Y^s_{out} e^{\iao z}/z,
\ena
the absorption probability $ {\G} $ can be calculated as\cite{TeukolskyPress}
\bea
1-\G = \vert {Y^s_{in}Y^{-s}_{in} \over Y^s_{out}Y^{-s}_{out}}\vert.
\ena
 From Eq.$(14)$, we find

\bea
Y_{out}^s&=&A_s4^p
(\iao )^{l+p} [1 - (-1)^{2p}(\iao)^{-(2l+1)}{\G(l+1-\iao) \over \G(-l-\iao)}]
,\nonumber\\
Y_{in}^s&=&A_s
(-\iao )^{l-p} \frac{\G (-l+s)}{\G (-l-s)}[1 +(-1)^{2l}(\iao)^{-(2l+1)}{\G(l+1-
\iao) \over \G(-l-\iao)}] ,
\ena
where
\bea
A_s=(-1)^{l+s}\frac{\G (1-s-\iao)}{\G(l+1-\iao)}.
\ena
By using Eqs.$(16)$ and $(17)$, we obtain
\bea
\G = {4\Delta_l\over {(1+\Delta_l)^2}},
\ena
where
\bea
\Delta_{l: {\rm integer}}=
\prod_{n=1}^{l}(1+\frac{n^2}{(a\omega)^2} ) \qquad (l \ge s \quad {\rm
for\quad bosons}) ,
\ena
\bea
  \Delta_{l:{\rm half-integer}}=\prod_{n=1}^{l+1/2}(1+\frac{(n-\frac{1}{2})^2}
  {(a\omega)^2})\qquad  (l \ge s \quad {\rm for \quad fermions}).
\ena
It is amazing to observe that the absorption probablility is independent of
the values
of spins and depends on spices (fermion or boson) and the total angular
momentum.
For spin 0 case, our result in Eqs.(19) and (20) agrees with the one derived by
Lohiya and
Panchapakesan\cite{RPone} and
the absorption probability vanishes for $l=0$. For $s=1,2$ cases, we have
non-vanishig
probabilities for all $l$  and our formula
are different from those by Lohiya and Panchapakesan[6].
In particular, we  found non-vanishing probability for $s=1/2$ in Eqs.(19) and
(21)
which is in contrast to the
result in Ref.\cite{RPtwo}  who argued that the absorption
probability is zero in their approximation.

For $s=0$, the functions $R_p$ is the same as $R_{-p}$ so that the trick
used to derive
$\G$ is not applicable. Lohiya and Panchapakesan\cite{RPone} derived  $\G$ and
showed that  the result agreed  with ours in Eqs.(19) and (20).
As for $s=1/2$, $R_{-p}$
is expressed in terms of $R_{p}$ so that another method using currents may
be useful.
For  this purpose, we can use the conserved current to  obtain the absorption
coefficients  as follows. We define
\bea
g_p = y^{(p+1)/2}(1-y)^{(s+1)}(1+y)^{-(s+p+1)}R_p
\ena
and then we find that
\bea
 W=g_{-p} ({d \over dy}g_p^* )-g_p^* ({d \over dy}g_{-p} )  = constant,
 \ena
which is the Wronskian.
For the solution in Eq.$(7)$ for  spin $1/2$  case,  we have
 \bea
g_{1/2} &=& y^{-{1 \over2}(\iao - {1 \over 2})}(1-y)^{l+1}F(l+{1 \over 2},l+1
- \iao,-\iao + {1 \over2};y),
\nonumber\\ g_{-1/2} &=& y^{-{1 \over2}(\iao - {3 \over 2})}(1-y)^{l+1}F(l+{3
\over 2},l+1 - \iao,-\iao + {3 \over2};y).
\ena
Since the functions $g_{1/2}$ and $g_{-1/2}$ are not independent, we can
rewrite the conservation of current in the following form;
\bea
W =y^{-1/2}(1-y)^{-1}[g_{1/2}^*g_{1/2} - {(l+{1\over2})^2 \over a^2\omega^2
+ {1 \over 4}}g_{-1/2}^*g_{-1/2}] = \frac 12 + \iao,
\ena
where the value of $W$ is fixed by evaluation near the horizon ($y \sim 1$).
 We evaluate $W$ by using the outgoing flux at the region $z<<1$ and
$a\omega z >>1$ and can obtain the absorption
coefficient.

We have used the analytic continuation of the angular momentum. But we
cannot justify the procedure of the analytic continuation whereas the
justification can be achieved for the evaluation of the absorption coefficients
of black holes\cite{MST}. At
present, we cannot obtain the asymptotic expansion without using the
analytic continuation. We hope that more rigorous approach can be done.

\end{document}